# Magnetoelectric coupling at the domain level in polycrystalline ErMnO$_3$

*J. Schultheiß*[1,2,*], *L. Puntigam*[1], *M. Winkler*[1], *S. Krohns*[1],
*D. Meier*[3], *H. Das*[4], *D. M. Evans*[1,5], *I. Kézsmárki*[1]

[1] Experimental Physics V, University of Augsburg, 86159 Augsburg, Germany
[2] Department of Mechanical Engineering, University of Canterbury, 8140 Christchurch, New Zealand
[3] Department of Materials Science and Engineering, Norwegian University of Science and Technology (NTNU), 7034 Trondheim, Norway
[4] Laboratory for Materials and Structures, Tokyo Institute of Technology, 4259 Nagatsuta, Midori-ku, Yokohama, Kanagawa 226-8503, Japan
[5] Department of Physics, University of Warwick, Coventry, CV4 7AL, United Kingdom
*corresponding author: jan.schultheiss@canterbury.ac.nz

**We explore the impact of a magnetic field on the ferroelectric domain pattern in polycrystalline hexagonal ErMnO$_3$ at cryogenic temperatures. Utilizing piezoelectric force microscopy measurements at 1.65 K, we observe modifications of the topologically protected ferroelectric domain structure induced by the magnetic field. These alterations likely result from strain induced by the magnetic field, facilitated by intergranular coupling in polycrystalline multiferroics. Our findings give insights into the interplay between electric and magnetic properties at the local scale and represent a so far unexplored pathway for manipulating topologically protected ferroelectric vortex patterns in hexagonal manganites.**

The combination of magnetic and ferroelectric order in a single component, a so-called multiferroic,[1] allows a broad range of scientifically and technologically interesting physical phenomena. Intriguing examples are the polarization reversal by a magnetic field,[2] mutual reinforcement of caloric effects,[3] and fascinating optical phenomena[4]. Many of these potential applications stem from the magnetoelectric coupling that links (anti)ferromagnetic and ferroelectric orders at the domain level.[5] While some general observations about this coupling can be made, such as the distinction between type II multiferroics, where magnetic and ferroelectric orders arise together, and type I multiferroics, where they can be independent, direct measurements are crucial for understanding the specific interactions and emergent coupling phenomena unique to each material.

A prominent example for a type I multiferroic is the family of the hexagonal (h) manganites, h-$R$MnO$_3$ ($R$=Sc, Y, In, and Dy-Lu).[6,7] In h-$R$MnO$_3$, the ferroelectric polarization emerges as a byproduct of a geometrically driven phase transformation linked to the tilting of the MnO$_5$ bipyramids at the Curie Temperature $T_C$>1200 K,[6] followed by an antiferromagnetic ordering of the Mn$^{3+}$ spins at the Néel temperature, $T_N$<120 K.[8] Examples of the coupling in h-$R$MnO$_3$ include the change of the ferroelectric polarization as a function of the applied magnetic field,[9,10] the flip of the magnetic spins with ferroelectric polarization reversal,[11] and magnetic phase diagram modification via an electric field.[12] Magnetoelectric coupling phenomena in the hexagonal manganites were related to a prominent magnetoelastic effect[13] and the structural shift of atomic positions of Mn$^{3+}$,[14] indirectly resulting in a coupling between the magnetic and ferroelectric order. An additional complication of the coupling in this family of materials is the presence of topologically protected structural vortexes, which are known to pin the ferroelectric/multiferroic domain pattern.[15] Practically, this means that an electric field applied to this system can grow/shrink the ferroelectric domains but not erase them completely.[16,17] As the correlation between the ferroelectric and magnetic orders emerges on the level of the domains[18] and domain walls,[19] studying the influence of a magnetic field on the ferroelectric domain structure via imaging techniques can provide valuable insights into magnetoelectric coupling phenomena in h-$R$MnO$_3$.

Here, we investigate the effect of magnetic fields on the ferroelectric domain structure of h-ErMnO$_3$ polycrystals using a combination of macroscopic permittivity measurements and nanoscale domain mapping. Permittivity measurements at 1.94 K indicate a change of the dielectric response under an applied magnetic field. Performing piezoresponse force microscopy (PFM) at 1.65 K, we observe that the ferroelectric domains, and topologically protected vortexes, can be altered by magnetic fields. Our finding provides a way to manipulate the ferroelectric domain structure in polycrystalline h-ErMnO$_3$.[20]

Polycrystals of h-ErMnO$_3$ are synthesized via a solid-state synthesis approach from Er$_2$O$_3$ (99.9 % purity, Alfa Aesar, Haverhill, MA, USA) and Mn$_2$O$_3$ (99,0% purity, Sigma-Aldrich, St. Lois, MO, USA) raw materials. Details on drying and ball-milling conditions are provided in ref. 21. The heat-treatment procedure to densify the powder into millimeter-sized samples is carried out at a temperature of 1450°C for 12 hrs. The magnetic-field dependent magnetization has been studied utilizing a SQUID magnetometer (Quantum Design MPMS 3, San Diego, CA, USA). The macroscopic dielectric response was obtained in a plate capacitor geometry with painted silver electrodes utilizing an Alpha Analyzer (Novocontrol, Montabaur, Germany) in a magnetic field ranging from 0 to 7 T, applied perpendicular to the electric field. For cooling and heating, a $^4$He-bath cryostat (CryoVac GmbH, Troisdorf, Germany) was used. Prior to PFM scans, the sample was lapped with a 9 μm-grained Al$_2$O$_3$ water suspensions (Logitech Ltd, Glasgow, UK) and polished using silica slurry (SF1 Polishing Fluid, Logitech AS; Glasgow, Scotland). To map the room-temperature piezoresponse, the sample was excited with an alternating voltage (40.13 kHz, 10 V peak-to-peak) using an electrically conductive platinum tip (Spark 150 Pt, Nu nano Ltd, Bristol, UK) on a NT-MDT Ntegra Prisma system (NT-MDT, Moscow Russia). Cryogenic PFM data was obtained on an attoAFM I (attocube systems AG, Haar, Germany) system with conductive diamond tips (CDT-FMR, Nanosensors, Neuchatel, Switzerland). A frequency of 55 kHz with 10 V peak-to-peak excitation voltage was utilized, while magnetic fields up to 5 T were applied perpendicular to the surface of the sample.

We begin our analysis by characterizing the crystal structure of our polycrystalline h-ErMnO$_3$ sample. An X-Ray diffraction (XRD) pattern, displaying the hexagonal space group symmetry $P6_3cm$, is shown in the inset of Fig. 1. We next probe the piezoelectric response of our polycrystalline h-ErMnO$_3$ at room temperature. A representative PFM scan is displayed in the inset in Fig. 1. The spatial resolution of $R\cos\vartheta$ (amplitude $R$ and phase $\vartheta$ of the piezoelectric response) allows to distinguish domains with antiparallel orientation of the ferroelectric polarization, while the grain boundaries separating grains of different crystallographic orientations are displayed by dashed white lines as explained in detail elsewhere.[21-24] The PFM image reveals a pronounced contrast, corresponding to the characteristic ferroelectric domain structure of polycrystalline h-ErMnO$_3$, featuring a mixture of vortex- and stripe like domains that forms at $T_c \approx$ 1420 K.[25,26]

To explore the low-temperature response of our polycrystalline h-ErMnO$_3$ sample, we measure the macroscopic dielectric permittivity as a function of temperature over 2-250 K for a range of frequencies from 1 to 10$^3$ Hz. As displayed in Figure 1, the dielectric permittivity continuously decreases with decreasing temperature, which was previously explained by the suppression of barrier layer contributions.[27,28] A feature in the dielectric data is shown at around 80 K, which corresponds to $T_N$ of h-ErMnO$_3$. At $T_N \approx$ 80 K, a second order

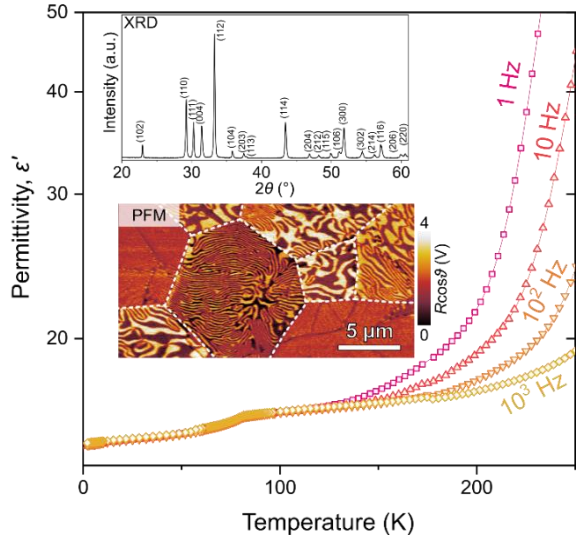

**Fig. 1.** Temperature-dependent dielectric permittivity, $\epsilon'$, of polycrystalline h-ErMnO$_3$ measured under different frequencies. The kink at about 80 K indicates the onset of antiferromagnetic order in h-ErMnO$_3$. Insets: The XRD pattern of crushed polycrystalline h-ErMnO$_3$ shows the hexagonal crystal structure with space group symmetry $P6_3cm$[31] and the out-of-plane PFM response of polycrystalline h-ErMnO$_3$. Dark and bright regions correspond to ±$P$ domains and dashed white lines mark the position of the grain boundaries.

phase transition from a paramagnetic (PM) to an antiferromagnetic (AFM) phase occurs.[29,30]

A simplified version of the temperature- and magnetic-field phase diagram of single crystalline h-ErMnO$_3$ is sketched in Figure 2a.[32] Figure 2b displays the magnetic-field-dependent magnetization of our samples measured at various temperatures between 2-80 K. At low temperatures, the magnetization exhibits a significant enhancement with a nonlinear dependence on the magnetic field. However, it's noteworthy that full saturation has not been attained within the investigated magnetic field range. This incomplete saturation can be attributed to the fact that the transition is influenced not only by the magnitude but also by the orientation of the field concerning the hexagonal axis of the individual grains. Consequently, the transition occurs over a broad range of magnetic fields related to the random crystallographic orientation of the individual grains in our polycrystalline material. Next, to investigate the influence of the magnetic field on the ferroelectric order of our polycrystalline h-ErMnO$_3$, we measure the magnetic field-dependent dielectric permittivity at 1.94 K for two frequencies ($10^4$ and $10^6$ Hz), as displayed in Figure 2c. At both frequencies, we find an anomaly in the dielectric response at around 0.8 T, which correlates to the AFM—FM transition indicated in Figure 2a, consistent with previous measurements on ErMnO$_3$ single crystals.[33,34] A change of the dielectric response under an applied magnetic field is often used as an experimental indication for the existence of a magnetoelectric interaction.[35-37] To exclude spurious effects resulting in a magnetic-field-induced signature in the dielectric response, e.g., magnetoresistance effects,[38] and to reveal the microscopic mechanism behind the observed feature, we next map the ferroelectric domain structure as a function of the magnetic field on the nanoscale.

We conduct this analysis by mapping the topography together with the amplitude and phase of the out-of-plane piezoelectric response at 1.65 K. We display the ferroelectric domain structure before (Figure 3a-e) and at a magnetic field of 5 T, which is applied perpendicular to the scan directions of the cantilever (Figure 3f-j). Note that because of the polycrystalline nature of the sample, the direction of the magnetic field depends on to the crystallographic orientation of the grain.[37] Domain structure schematics, which are reconstructed from the PFM amplitude and phase data, are presented in Figure 3d and I, showing the domain structure at 0 T and 5 T, respectively. The schematics indicate that the vortex core has moved after application of the applied magnetic field, leaving locally a purely stripe-like domain structure behind. Such movement is unexpected as ferroelectric domain structures are typically more flexible at higher temperatures, and even at high temperatures, the conjugate field to the polar order is typically unable to move the vortex cores.

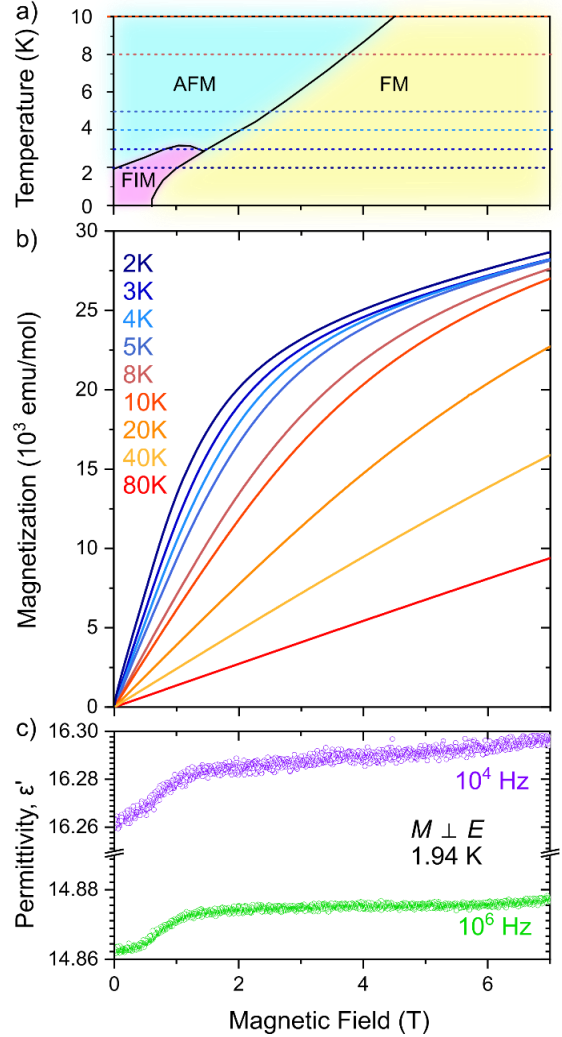

**Fig. 2.** a) Simplified magnetic field versus temperature phase diagram for an h-ErMnO$_3$ single crystal, redrawn from ref. 32. The antiferromagnetic (AFM), the ferrimagnetic (FIM), and ferromagnetic (FM) phases are indicated. B) The magnetic-field dependence of the magnetization of h-ErMnO$_3$ polycrystals is measured at different temperatures (as indicated in panel a) by dashed lines of corresponding colors). c) Magnetic field-dependent dielectric permittivity of polycrystalline h-ErMnO$_3$ measured at $10^4$ and $10^6$ Hz at a temperature of $T$=1.94 K.

We suggest two possible mechanisms to explain this result. Either, a direct magnetic field effect on the collective magnetic moment at the domain walls,[19] or a strain-induction coupling through the magnetoelastic effect of h-ErMnO$_3$[13]. To distinguish these, we take line profiles in Figure 3e and j that show the change of the domain structure is driven by the vortex core. Moreover, since the domain structure in single crystals of h-ErMnO$_3$ was found to be independent of the applied magnetic field up to magnetic fields of 4 T at 2.8 K,[20] we suggest that the observed changes are a consequence of the polycrystalline nature of the sample. It is established that vortex cores in h-$R$MnO$_3$ interact with strain fields, and a strain-induced movement of the vortex cores was theoretically[39] and experimentally[22,40,41] demonstrated for temperatures around $T_c$. In our case, strain may arise from substantial magnetoelastic coupling and the magnetostrictive response in the hexagonal manganites, facilitated by the clamping of the individual grains.

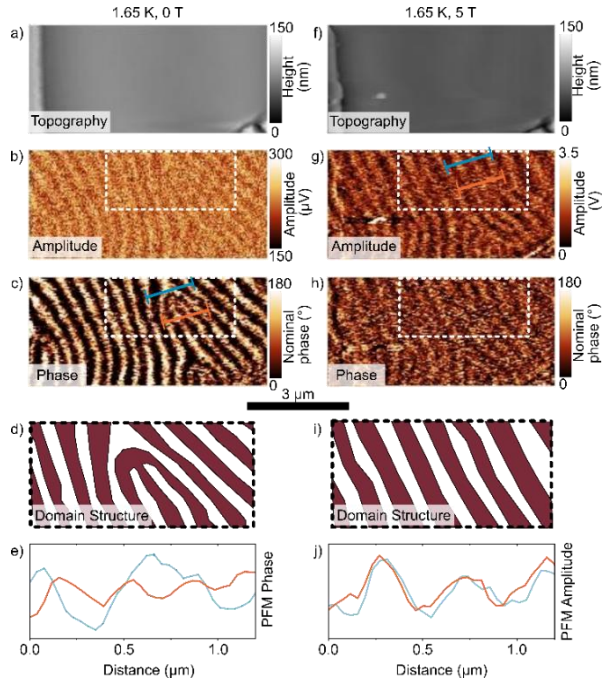
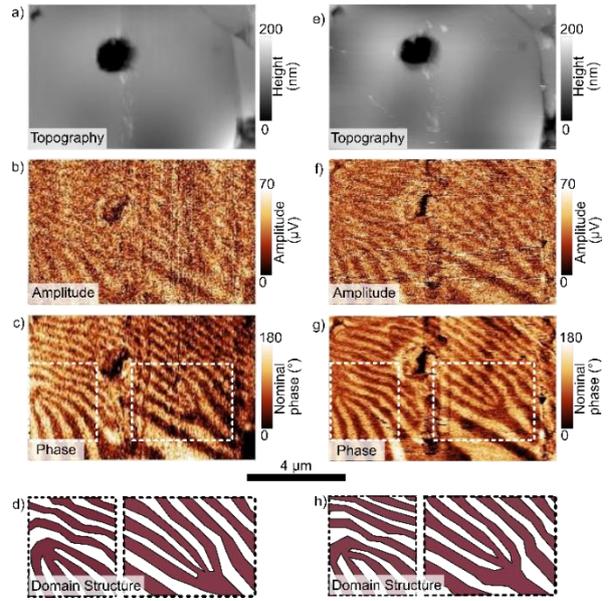

**Fig. 3:** Data obtained at 1.65 K without a)-e) and with e)-j) applied 5 T magnetic field are shown. Topography images are presented in a) and f). The corresponding PFM amplitude and phase are depicted in b) and c) for 0 T, while the influence of a magnetic field of 5 T is illustrated in g) and h), respectively. The schematic drawing of the domain structure in d) and i) indicates a modified ferroelectric domain structure after application of the magnetic field. Line profiles extracted from experimental data, showcased in e) and j), highlight the modification of the ferroelectric domain structure with a magnetic field applied.

The impact of the magnetic field on the domain structure appears to be highly stochastic, with observable changes occurring inconsistently across different areas of the sample. A representative area where the ferroelectric domain structure is not impacted by the magnetic field is displayed in Figure 4. Figure 4a-c illustrates PFM measurements before applying the magnetic field, while Figure 4e-g shows the corresponding results recorded under a magnetic field of 5 T. The PFM results are illustrated by the schematic domain structure drawn in Figure 4d and h. We do not resolve any field-induced change in the domain structure at a magnetic field of 5 T for this specific area. The observed stochastic nature of the magnetic-field induced domain wall movement may be related to different absolute strain values due to the random orientation of the grains in combination with spatially different types of vortex cores and coupling behavior. Further, the mobility of ferroelectric domain walls at 1.65 K is expected to be low, [42,43] related to thermal activation and spatially different pinning potentials.

In summary, our study demonstrates that the ferroelectric domain structure of polycrystalline h-ErMnO$_3$ can be manipulated using a magnetic field of 5 T at around 2 K, as evidenced by a combination of macroscopic dielectric and atomic-force microscopy measurements. We observe that this effect is not universal and appears highly stochastic in some areas of the sample. Further investigation is required to understand how the interplay of grain orientations with ferroelectric domains influences the field-induced mobility of vortices, necessitating studies with larger statistics. Unlike an electric field, which causes the contraction of domains into meandering bands,[16,17,44] our findings suggest that the predominant mechanism involves the interaction between a magnetic field and a vortex core resulting in a predominant stripe-like domain structure. The interaction between the magnetic field and the domain structure is indirect and possibly occurs through a magnetic field-induced elastic strain.

**Fig. 4.** Data obtained at 1.65 K without a)-d) and with e)-h) an applied magnetic field are presented. Topography images obtained at the same position are shown in a) and e). The corresponding PFM amplitude and phase are depicted in b) and c) for 0 T, while the influence of a magnetic field of 5 T is illustrated in f) and g). The schematic drawing of the domain structure in d) and h) indicates that the applied magnetic field has no influence on the ferroelectric domain structure in this area.

Consequently, we attribute the observed effect to the polycrystalline nature of h-ErMnO$_3$ and extend intergranular coupling, which already determines the switching mechanism in polycrystalline ferroelectric/ferroelastic materials[45-47] towards multiferroics in their polycrystalline form. Moreover, the observed indirect magnetoelectric coupling suggests the same unexpected properties may be present in other type-I polycrystalline multiferroics, hosting vortex cores, such as hexagonal indates,[48] ferrites,[49] or vanadates[50] providing a rich platform for studying the influence of different magnetic sublattices on the interaction.

### ACKNOWLEDGEMENTS
D. Vieweg is acknowledged for performing the squid measurements. N. Domingo, G. Catalán, and Th. Lottermoser are acknowledged for helpful discussions. J.S. acknowledges financial support from the Alexander von Humboldt Foundation through a Feodor-Lynen research fellowship, the German Academic Exchange Service (DAAD) for a Post-Doctoral Fellowship (Short-term program), and NTNU Nano through the NTNU Nano Impact fund. D.M.E. acknowledges and thanks the Deutsche Forschungsgemeinschaft for financial support via an individual fellowship number (EV 305/1-1) and the Engineering and Physical Sciences Research Council (EP/T027207/1). D.M. thanks NTNU for support through the Onsager Fellowship Program, the outstanding Academic Fellow Program, and acknowledges funding from the European Research Council (ERC) under the European Union's Horizon 2020 Research and Innovation Program (Grant Agreement No. 863691).

### DATA AVAILABILIY
The data that supports the findings of this study are available from the corresponding author upon reasonable request.